\documentclass[prl,twocolumn,amssymb,amsmath]{revtex4-1}
\usepackage{graphicx}
\usepackage{epsfig,color}
\usepackage{amsmath,amssymb}

\begin{document}

\title{Simulating Compact Quantum Electrodynamics with ultracold atoms:\\
Probing confinement and nonperturbative effects}

\date{\today}

\author{Erez Zohar$^1$, J. Ignacio Cirac$^2$, Benni Reznik$^1$}

\begin{abstract}
\begin{center}
$^1$School of Physics and Astronomy, Raymond and Beverly Sackler
Faculty of Exact Sciences, Tel Aviv University, Tel-Aviv 69978, Israel.

$^2$Max-Planck-Institut f\"ur Quantenoptik, Hans-Kopfermann-Stra\ss e 1, 85748 Garching, Germany.
\end{center}
Recently, there has been much interest in simulating quantum field theory effects of matter and gauge fields. In a recent work, a method for simulating compact Quantum Electrodynamics (cQED) using Bose-Einstein condensates has been suggested. We suggest an alternative approach, which relies on single atoms in an optical lattice, carrying $2l+1$ internal levels, which converges rapidly to cQED as $l$ increases. That enables the simulation of cQED in 2+1 dimensions in both the weak and the strong coupling regimes, hence allowing to probe confinement as well as other nonperturbative effects of the theory. We provide an explicit construction for the case $l=1$ which is sufficient for simulating the effect of confinement between two external static charges.
\end{abstract}

\maketitle

Dynamic Gauge theories are in the core of the standard model of particle physics, playing the role of the force carriers among the matter fields, and therefore are of particular significance.
It was shown, using lattice gauge models and other methods that such gauge theories exhibit the peculiar phenomenon of confinement of charges which is related to non-perturbative effects due to non linear
interactions in the theory\cite{Wilson,KogutSusskind,Polyakov}.
Such lattice gauge theories are believed to have a non-trivial phase structure. The simplest such theory is compact QED (cQED) - a U(1) lattice gauge theory, which is believed to manifest, in 3+1 dimensions, a phase transition between the confining phase (for large coupling constant $g$) and the nonconfining Coulomb phase (for small coupling), while in 2+1 dimensions it was shown that the theory confines also in the weak coupling regime because of nonperturbative effects \cite{Wilson,KogutSusskind,Polyakov,BanksMyersonKogut,DrellQuinnSvetitskyWeinstein,BenMenahem}.
In non-abelian Yang-Mills theories, it is believed that confinement holds for all values of the coupling constants.

Recently there has been much interest in quantum simulations of quantum field theories by utilizing methods in ultracold atoms
and other systems \cite{CiracZoller}.   Models have been suggested for simulating dynamical matter fields\cite{Cirac2010,Bermudez2010,Boada2011}, and exotic phenomena manifested by such fields have been discussed\cite{Retzker2005, Horstmann2010,Boada2012}.  However, less progress has been achieved for dynamic abelian and non-abelian gauge theories. Dynamic gauge theories involving spin-half states have been discussed in \cite{Pyrochlore,Rydberg}.
Since the electric fields in such models can obtain only two values, such models are unable to manifest the effect of electric flux-tubes (but rather of different "strings").
Coloumb phase simulations have been suggested with molecular states \cite{RingExchange}, and BECs in optical lattices \cite{ArtificialPhotons}.

In a recent work \cite{Zohar2011}, we have obtained, using Bose Einstein condensates in an optical lattice, an effective theory of a dynamic U(1) gauge theory, manifesting confinement of external static charges, with observable electric flux-tubes.
 In the present paper we suggest an alternative approach for simulating gauge theories in terms of a Spin-Gauge Hamiltonian $H_{l}$ (defined in equation \ref{SG}), which describes
  interacting single atoms with internal levels playing the role of angular momentum multiplet $-l\leq m \leq l $, instead of BECs as \cite{Zohar2011}. We will show that for large values of $l$  the
Spin-Gauge Hamiltonian rapidly converges to the standard abelian Kogut-Susskind model \cite{KogutSusskind,KogutLattice} of cQED for both the weak and strong coupling regimes.
  Hence this model is able to simulate the effect of confinement as well as non-perturbative effects in the weak coupling regime, which give rise to it in 2+1D. As a first step in realizing the models $H_{l}$ we shall propose a method for constructing the case $l=1$ which
  is sufficient for observing the effect of confinement of static charges in the strong regime. First, using the methods of \cite{Mazza2010}, we construct an effective "generalized XXZ" model. Then, adapting the ideas of \cite{Zohar2011}, the gauge invariance is introduced to the system with a second effective calculation.
  The single atoms implementation might be easier experimentally than the BEC approach, as it does not rely on the overlaps of local single-particle wavefunctions and thus the energy scales in the Hamiltonian may be larger. Moreover, only three atomic species are required, and they can populate every link, unlike in the BEC model.

\begin{figure}[t]
\includegraphics[scale=0.7]{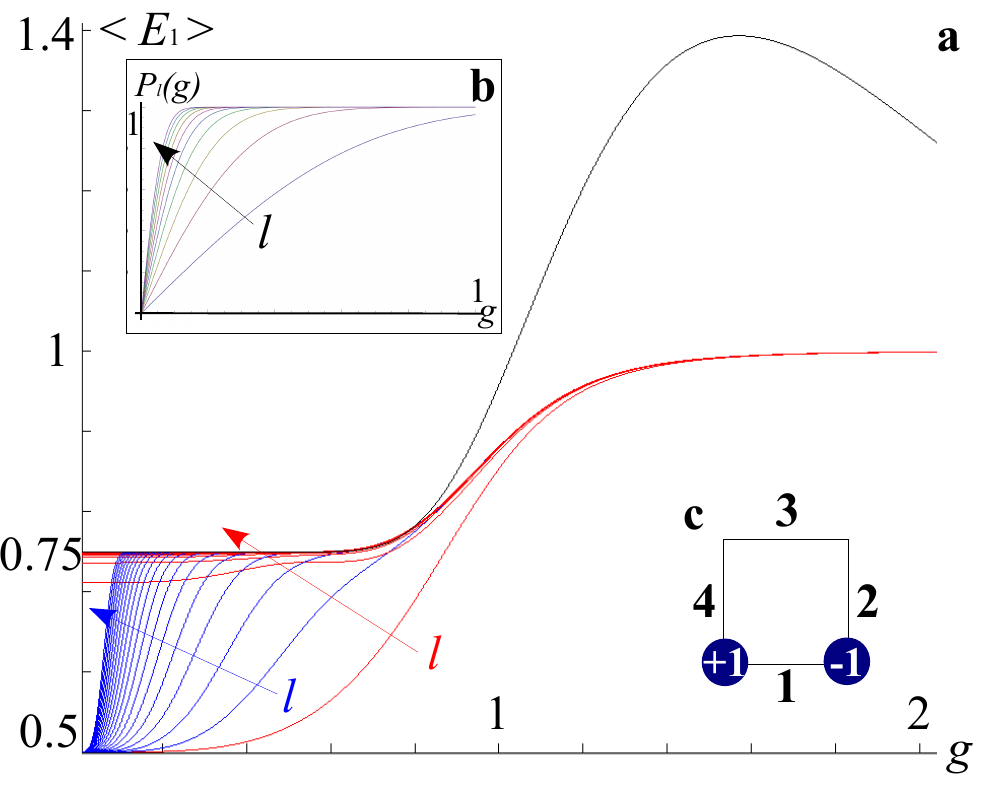}
\caption{\emph{Single Plaquette Plots.} (a) Graphs of $\left\langle E_1 \left(g \right) \right\rangle$. Black - the calculation of \cite{DrellQuinnSvetitskyWeinstein} - small coupling approximation for regular abelian Kogut-Susskind Theory. Blue - exact calculation for the truncated theory, for for $l=1,...,20$. Red - Exact calculation for the Spin-Gauge theory, for $l=1,...,20$. It can be seen that the curves start to merge for a small $g$ and $l$. The value 1 refers to the flux-tube and $\frac{3}{4}$ to the longitudal part. (b) Graphs of $P_l \left(g\right)$, for $l=1,...,20$. It can be seen that even for small values of $l$, $P_l \left(g\right)$ approaches 1 for a small $g$. (c) The one plaquette system we use in the demonstration. }
\label{exact}

\end{figure}

 Let us consider a 2D square lattice with single atoms which carry 2l+1 internal states, located on the
 links and described by the Spin-Gauge (SG) Hamiltonian:
\begin{widetext}
\begin{equation}
H_{l}=\frac{g^{2}}{2}\underset{\mathbf{n},k}{\sum}\left(L_{z\mathbf{n}}^{k}\right)^{2}-\frac{1}{2g^{2}\left(l\left(l+1\right)\right)^{2}}\underset{\mathbf{n}}{\sum}\left(L_{+,\mathbf{n}}^{1}L_{+,\mathbf{n+\hat{1}}}^{2}L_{-,\mathbf{n+\hat{2}}}^{1}L_{-,\mathbf{n}}^{2}+h.c.\right)
\label{SG}
\end{equation}
\end{widetext}
 where $\mathbf{n}$ are the lattice's vertices, $k=1,2$ are the lattice directions, whose corresponding unit vectors are $\mathbf{\hat 1},\mathbf{\hat 2}$. For example, $L_{z\mathbf{n}}^{k}$ is the z component
of the spin on the link emanating from the vertex $\mathbf{n}$ in the $k$th direction (The generalization to a 3D lattice is straightforward). $g$ is a constant.
This should be compared to the abelian Kogut-Susskind Hamiltonian \cite{KogutSusskind,KogutLattice}
$H_{KS} = \frac{g^{2}}{2}\underset{\mathbf{n},k}{\sum}\left(E_{\mathbf{n}}^{k}\right)^{2}-\frac{1}{g^{2}}\underset{\mathbf{n}}{\sum}\cos\left(\phi_{\mathbf{n}}^{1}+\phi_{\mathbf{n + \hat 1}}^{2}- \phi_{\mathbf{n + \hat 2}}^{1} - \phi_{\mathbf{n}}^{2}\right)$.
Unlike the Kogut-Susskind Hamiltonian, in our case we are dealing with 3D angular momentum operators.
 Nevertheless, for large values of $l$,
 the first quadratic term in the Hamiltonian conincides with the electric part of the abelian Kogut-Susskind Hamiltonian , with the z components of the angular momentum playing the role of an electric field $E$, and the second, quartic part with the magnetic part of the Kogut-Susskind Hamiltonian.
 This can be seen qualitatively when considering the matrix elements of $\left|m\right| \ll l$, for which
 $\frac{L_{\pm}}{\sqrt{l\left(l+1\right)}}\left|l,m\right\rangle \approx \left|l, m \pm 1\right\rangle$, similarly to
 $e^{\pm i \phi}\left|m\right\rangle = \left|m \pm 1\right\rangle$ in the Kogut-Susskind model. We shall test this equivalence quantitatively for the case of a single plaquette (figure \ref{exact}).

 For every $l$, the Spin-Gauge Hamiltonian manifests a local U(1) gauge symmetry, that is generated by the local operators (defined on the vertices of the lattice) $G_{\mathbf{n}}=\underset{k}{\sum}\triangle_{k}L_{z\mathbf{n}}^{k}$ (where $\triangle_{k}f_{\mathbf{n}}=f_{\mathbf{n+\hat k}}-f_{\mathbf{n}}$) which commute with the Hamiltonian: for a given vertex, $G_{\mathbf{n}}$ trivially commutes with all the plaquettes which do not contain $\mathbf{n}$. As for the other four plaquettes, the commutation relation is zero, since $\left[L_z,L_{\pm}\right]=\pm L_{\pm}$. For example,
$\left[G_{\mathbf{n}},L_{+,\mathbf{n}}^{1}L_{+,\mathbf{n+\hat{1}}}^{2}L_{-,\mathbf{n+\hat{2}}}^{1}L_{-,\mathbf{n}}^{2}\right] =
\left[L_{z \mathbf{n}}^1+L_{z \mathbf{n}}^2,L_{+,\mathbf{n}}^{1}L_{+,\mathbf{n+\hat{1}}}^{2}L_{-,\mathbf{n+\hat{2}}}^{1}L_{-,\mathbf{n}}^{2}\right] =
L_{+,\mathbf{n}}^{1}L_{+,\mathbf{n+\hat{1}}}^{2}L_{-,\mathbf{n+\hat{2}}}^{1}L_{-,\mathbf{n}}^{2}-L_{+,\mathbf{n}}^{1}L_{+,\mathbf{n+\hat{1}}}^{2}L_{-,\mathbf{n+\hat{2}}}^{1}L_{-,\mathbf{n}}^{2}=0$.
Static external charges $\left|\left\{ Q_{\mathbf{n}}\right\} \right\rangle $ are introduced to the system by fixing a subspace by the constraint
$G_{\mathbf{n}} \left|\left\{ Q_{\mathbf{n}}\right\} \right\rangle = Q_{\mathbf{n}} \left|\left\{ Q_{\mathbf{n}}\right\} \right\rangle$.

In order to have something useful for simulations, we would like to have that $H_{l} \rightarrow H_{KS}$ for large $l$s, sufficiently fast. Thus we shall consider a comparison between the Spin-Gauge Hamiltonian with a constant $l$ and a truncated version of the Kogut-Susskind Hamiltonian with $-l \leq E \leq l$.
 It is straightforward to see, using perturbation theory in $g^{-1}$, that in the strong limit of the Hamiltonian ($g \gg 1$) the ground states of the Spin-Gauge and Kogut-Susskind Hamiltonians coincide up to a certain order in the perturbative expansion, depending on $l$ and the charge distribution. On the other hand, in the weak coupling limit, we shall examine the effect of truncation in a nonperturbative manner, for a single plaquette system.

 \emph{Case of a single plaquette.} Consider a single plaquette with two opposite unit static charges in the lower vertices (see figure \ref{exact}c). Using the gauge invariance and Gauss's law, a possible gauge-invariant basis of states is $\left|m\right\rangle \equiv \left|m,m-1,1-m,1-m\right\rangle$ ($m=-l+1,...,l$) where these are the eigenvalues of the electric field on each link, from the lower one, counter-clockwise.
  Relying upon the results of Drell et. al. in \cite{DrellQuinnSvetitskyWeinstein}, the ground state of this system, for weak coupling, is given by a Bloch-like wavefunction in the tight-binding limit
 $\chi\left(\theta\right) = \overset{\infty}{\underset{n=-\infty}{\sum}}e^{\frac {in\pi}{2}} e^{- \frac{1}{4g^{2}}\left( \theta - 2 \pi n \right)^2}$, where $\theta$ is the magnetic field in the plaquette.
On the link connecting the charges, $\left\langle E_1 \right\rangle = \frac{3}{4} + \frac{\pi}{g^4}\left(\frac{\pi^2 -4}{2}\right)e^{- \frac{\pi^2}{2g^{2}}}$. This can be understood as the contributions of two parts: One is the trivial contribution of the longitudal, classical, static Coulomb field ($\frac{3}{4}$). The second part is much more interesting: it is nonanalytic in $g=0$ and therefore it is nonperturbative in $g$. It is this type of mechanism which is responsible for confinement in the weak regime in large 2D spatial lattices. In figure \ref{exact}a, $\left\langle E_1 \right\rangle$ is plotted as a function of $g$ - both for the truncated Kogut-Susskind Hamiltonian and the Spin-Gauge Hamiltonian for several values of $l$, and the analytic result of \cite{DrellQuinnSvetitskyWeinstein} as well. For $g<1$, the truncated and Spin-Gauge results conincide with each other and with the analytical approximation for $l \approx 2,3$. For large values of $g$, the flux-tube value of 1 is reached already for $l=1$.

In order to understand the effect of truncation, we expand this state in the $m$ basis -
$\left| \chi \right\rangle = \overset{\infty}{\underset{m=-\infty}{\sum}} a_{m}\left|m\right\rangle = \overset{\infty}{\underset{m=-\infty}{\sum}} e^{-g^2\left(m-\frac{3}{4}\right)^2}\left|m\right\rangle$,
and take only $-l+1 \leq m \leq l$. As a measure of the accuracy of truncation we calculate the probablity to be in the truncated subspace: $P_l \left(g\right) = \frac{1}{\left\langle\chi | \chi \right\rangle} \overset{l}{\underset{m=-l+1}{\sum}} \left|a_{m}\right|^2$
as one can see in figure \ref{exact}b, this function approaches $1$ quickly even for small finite $l$s, which means that the truncated theory still shows the same effect for small $g$s.

\begin{figure}[t]
\includegraphics[scale=0.45]{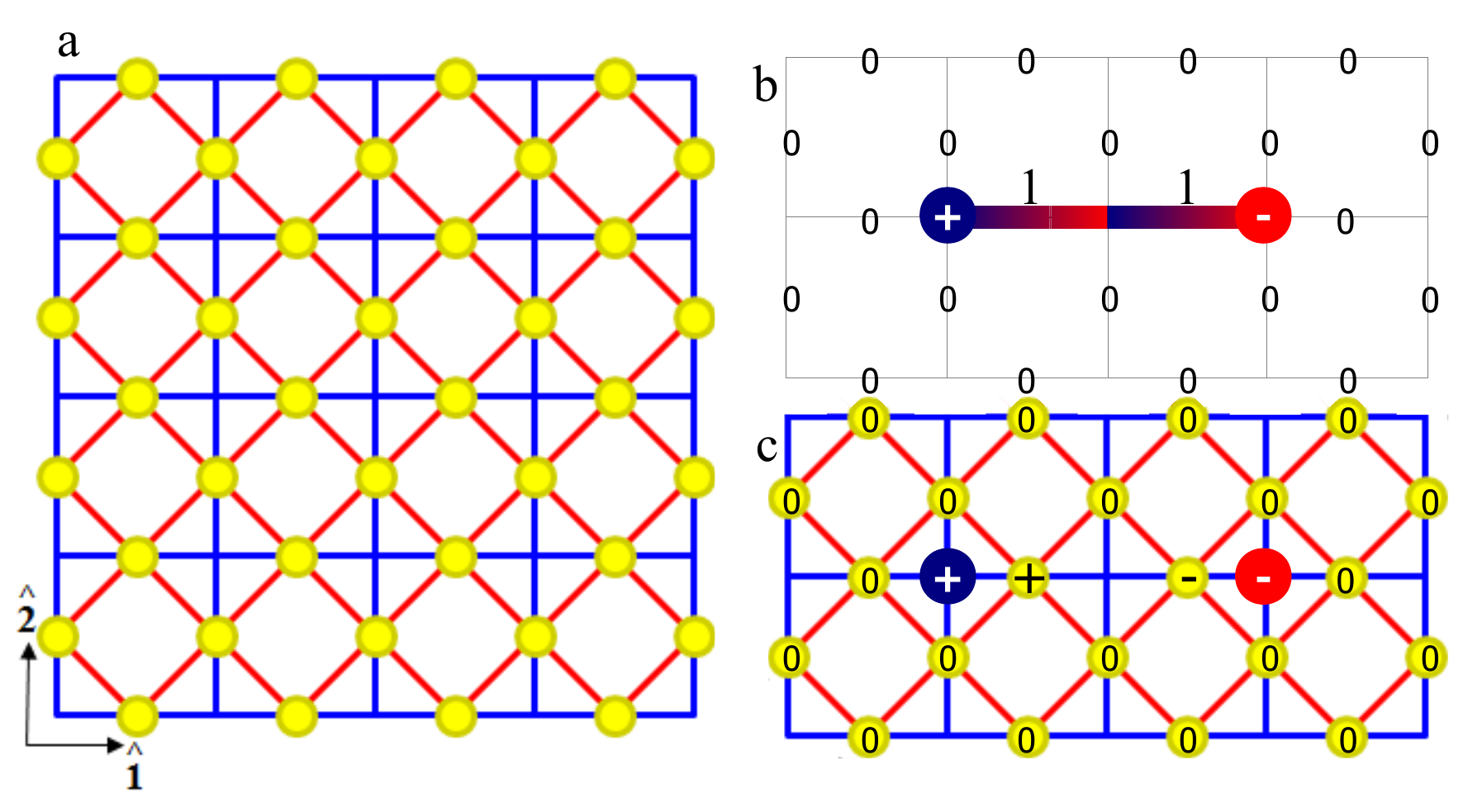}
\caption{(a) The simulation lattice. As explained in the text, the atoms (yellow circles) are
aligned along the links of a 2d square lattice, with basis vectors $\mathbf{\hat 1}$ and $\mathbf{\hat 2}$.
On the blue lines the tunneling rates between neighboring atoms are $t_{\alpha}^{s}$ (in Hamiltonian (2)),
and on the red lines - $t_{\alpha}^{d}$. In Hamiltonian (3), $z-z$ interaction are between nearest neighbors
along both red and blue lines, and $x-x$, $y-y$ interactions are only along the red ones.
(b) The flux tube (zeroth order in perturbative expansion, in the strong limit of the Kogut-Susskind Hamiltonian) connecting two opposite charges.
(c) The flux tube in the language of our simulating system.  }
\label{lattice}
\end{figure}

\emph{Simulating the $l=1$ Spin-Gauge Hamiltonian.} Let us consider a 2d square optical lattice\cite{Lewenstein2012}, whose minima conincide with the links of the square lattice of the Spin-Gauge Hamiltonian (the generalization to 3d is straightforward).
Each minimum is populated by a single atom of three different atomic levels, forming a $l=1$ spinor (see figure \ref{lattice}a). We use it to develop an effective theory \cite{soliverez1969}, which manifests confining flux tubes similar to the ones in cQED. We first turn to the derivation of an effective generalized spin-1 XXZ Hamiltonian on this lattice.

The Hamiltonian describing the dynamics of three atomic species $\alpha \in \{ +,0,- \}$ on such an optical lattice contains same-species tunneling terms along the
$\mathbf{\hat 1},\mathbf{\hat 2}$ directions and the diagonal directions as well, and on-site spin and number dependent terms. A pictorial representation of the interactions  can be found in figure \ref{lattice}a, and a more deatiled description of the lattice structure and the optical potential can be found in the supplementary material. Generalizing from \cite{Mazza2010} we obtain the Hamiltonian:
\begin{widetext}
\begin{multline}
H = -\underset{\mathbf{n},\alpha}{\sum}
\left(t_{\alpha}^{s}
\left(a_{\mathbf{n},\alpha}^{\dagger}a_{\mathbf{n}+\mathbf{\hat 1},\alpha} + b_{\mathbf{n},\alpha}^{\dagger}b_{\mathbf{n}+\mathbf{\hat 2},\alpha}
\right) +
t_{\alpha}^{d}
\left(a_{\mathbf{n},\alpha}^{\dagger}b_{\mathbf{n},\alpha} + a_{\mathbf{n},\alpha}^{\dagger}b_{\mathbf{n}+\mathbf{\hat 1},\alpha}  +
b_{\mathbf{n},\alpha}^{\dagger}a_{\mathbf{n}+\mathbf{\hat 2},\alpha} +
a_{\mathbf{n}+\mathbf{\hat 2},\alpha}^{\dagger}b_{\mathbf{n}+\mathbf{\hat 1},\alpha}
\right)
+ H.C. \right)
+ \\
+ \underset{\mathbf{n},k,\alpha}{\sum}\Delta_{\mathbf{n},\alpha}^{k}N_{\mathbf{n},\alpha}^{k}
+ \frac{U_{0}}{2}\underset{\mathbf{n},k}{\sum}N_{\mathbf{n}}^{k}\left(N_{\mathbf{n}}^{k}-1\right)
+ \frac{U_{2}}{2}\underset{\mathbf{n},k}{\sum}\left(\vec{S_{\mathbf{n}}^k}^{2}-2N_{\mathbf{n}}^{k}\right)
\end{multline}
\end{widetext}
where for horizontal links the annihilation operators are $a_{\mathbf{n},\alpha}$, and the number operators - $N_{\mathbf{n},\alpha}^{1}=a_{\mathbf{n},\alpha}^{\dagger}a_{\mathbf{n},\alpha}$, and for vertical links -$b_{\mathbf{n},\alpha}$ and $N_{\mathbf{n},\alpha}^{2}=b_{\mathbf{n},\alpha}^{\dagger}b_{\mathbf{n},\alpha}$;
$N_{\mathbf{n}}^{k} =\underset{\mathbf{\alpha}}{\sum}N_{\mathbf{n},\alpha}^{k}$ and
$\vec{S_{\mathbf{n}}^k}$ is the total on-site spin (see \cite{Mazza2010}).

We set the parameters $\Delta_{\mathbf{n},\pm}^{k}=\frac{\delta}{2} + 2\lambda + \mu \mp 2\lambda\left(q_{\mathbf{n}}+q_{\mathbf{n+\hat k}}\right)$ , $\Delta_{0}=0$,
with $\delta \ll U_{0},U_{2}$ , $\mu \ll \lambda \ll U_{0},U_{2}$, and $q_{\mathbf{n}}$ are integer C-numbers which will be later related to the static charges. We also introduce a new variable $z$, satisfying $U_{2}=zU_{0}$.

\emph{Derivation of a generalized XXZ Hamiltonian.} The first effective calculation, which leads to a generalized XXZ model, is similar to the one in \cite{Mazza2010}.
Since the $U$ local terms are much larger than the others, it is reasonable to obtain, perturbatively,
an effective Hamiltonian around them \cite{soliverez1969}. Unlike in \cite{Mazza2010}, we do not include the small local terms
$\underset{\mathbf{n},k,\alpha}{\sum}\Delta_{\mathbf{n},\alpha}^{k}N_{\mathbf{n},\alpha}^{k}$ in the constraining part of the Hamiltonian, but rather treat them as the first order contribution to the
effective Hamiltonian. We divide that into two parts: one is the $\lambda,\mu,q_{\mathbf{n}}$ dependent part, which we put aside at the moment; The other $\delta$ dependent part will be used to construct "two-site" connected local diagonal Hamiltonians: the energy contribution of this part from each link will be equally distributed among the six connections it has with other links (to whom it is connected by tunneling).

The tunneling rates are chosen to be real: $t_{+}^{s}=t_{-}^{s}=t_{+}^{d}=t_{-}^{d}=\frac{1}{4}\sqrt{\frac{U_{0}\left(24\lambda-5\epsilon\right)\left(\epsilon+24\lambda\right)}{6\epsilon}}$,
$t_{0}^{s}=0$,
$t_{0}^{d}=\Omega\sqrt{\frac{3U_{0}\left(24\lambda-5\epsilon\right)}{2\epsilon\left(\epsilon+24\lambda\right)}}$,
where $\frac{\lambda^{2}}{U_{0}}$,
$\frac{\Omega^{2}}{U_{0}}\ll\epsilon\ll\Omega\ll\lambda$
and $t_{0}^{d}\ll t_{\pm}^{d}$.
Finally we set $\delta=-12\lambda-\epsilon$ and $z=\frac{1}{4}-\frac{6\lambda}{\epsilon}$. Then we can apply the effective calculation as in \cite{Mazza2010}.
Collecting these resutls with the $\lambda,\mu,q_{\mathbf{n}}$ dependent part, we get,
up to a constant energy, the first effective Hamiltonian,
\begin{widetext}
\begin{equation}
H_{eff}^{(1)}= 2 \lambda \underset{str+diag}{\sum} L_{z,\mathbf{n}}L_{z,\mathbf{n'}} + \Omega \underset{diag}{\sum} \left(L_{x,\mathbf{n}}L_{x,\mathbf{n'}}+L_{y,\mathbf{n}}L_{y,\mathbf{n'}}\right) +\underset{\mathbf{n},k}{\sum}
\left(\left(2\lambda+\mu\right)\left(L_{z,\mathbf{n}}^{k}\right)^{2}
-2\lambda\left(q_{\mathbf{n}}+q_{\mathbf{n + \hat k}}\right)L_{z,\mathbf{n}}^{k} \right) + O(\epsilon)
\end{equation}
\end{widetext}
where the $z-z$ interactions are between links who share a vertex and the $x-x$,$y-y$ only between links who share both a vertex and a plaquette (see figure \ref{lattice}a). This is an example of a frustrated XXZ model Hamiltonian, which is of interest of its own. One can easily check that the scale hierarchy is not violated.

\begin{figure}[!h]
\begin{center}
\includegraphics[scale=0.4]{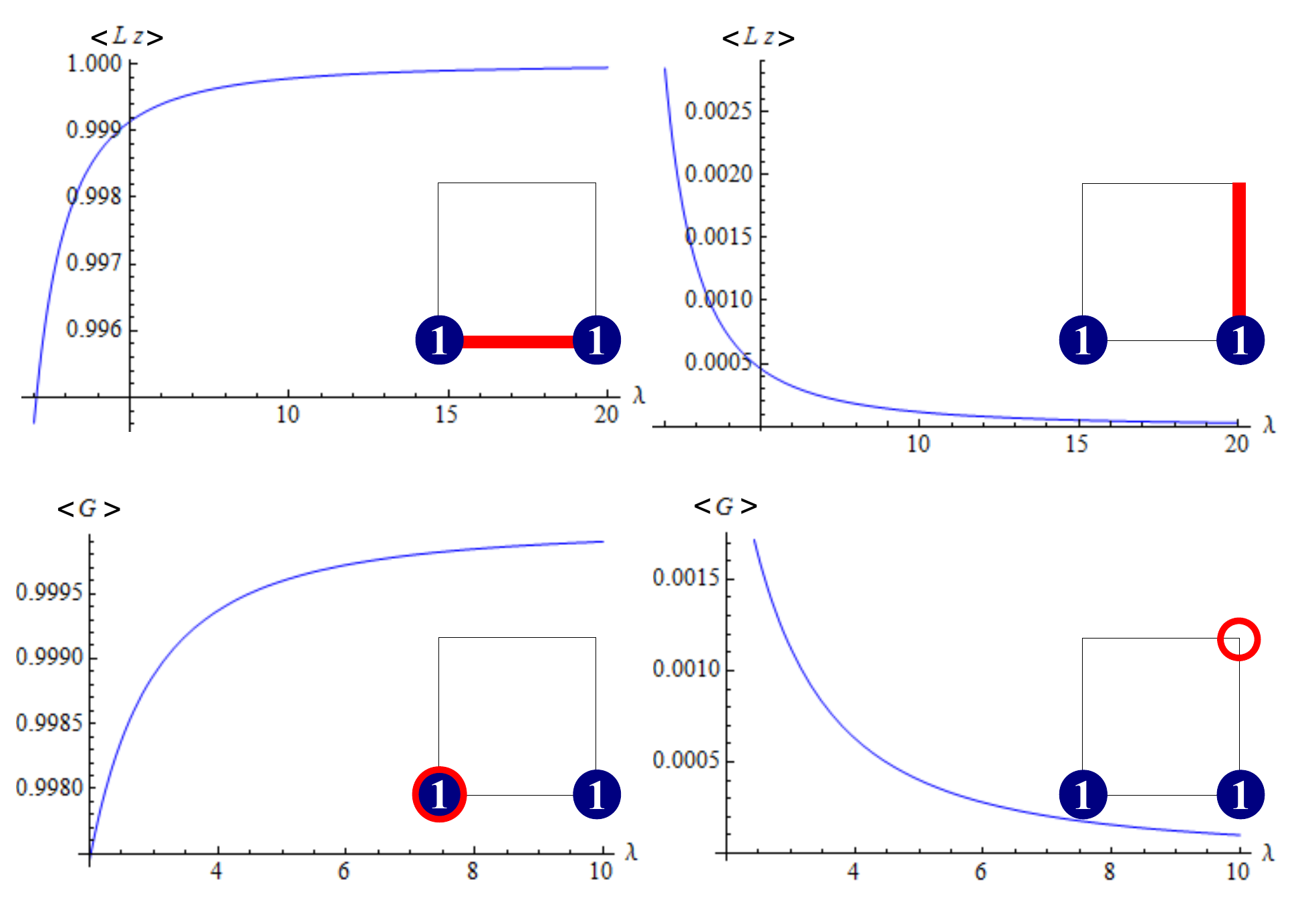}
\caption{An example of the emergence of gauge invariance in the second effective calculations for a
single plaquette system ($l$=1). Two unit charges are located in the lower vertices. The parameters are chosen such that for a large $\lambda$ the system will be in the strong limit: $\mu=1$, $\Omega=0.1$.
The expectation values of $L_z$ along two links, as well as of the gauge transformation generators $G$ at two vertices are plotted as a function of $\lambda$ for the ground state. It can be seen that as $\lambda$ grows, $G$ converge to the local charges and hence Gauss's law and gauge invariance are introduced, and $L_z$ approach the values of the lowest flux-tube state in the strong limit. Note the discussion on the signs in the text.
}
\label{eff}
\end{center}
\end{figure}

\emph{Imposing gauge invariance on the system}. In the second step, we employ the method of \cite{Zohar2011}. Define $c_{\mathbf{n}} \equiv \frac{1}{\sqrt{2}} L_{- \mathbf{n}}^{1}$, $d_{\mathbf{n}} \equiv \frac{1}{\sqrt{2}} L_{- \mathbf{n}}^{2}$, $G_{\mathbf{n}} \equiv   L_{z,\mathbf{n}}^1+L_{z,\mathbf{n}}^2+L_{z,\mathbf{n-\hat1}}^1+L_{z,\mathbf{n-\hat2}}^2$, and rewrite the effective Hamiltonian as
\begin{widetext}
\begin{equation}
H_{eff}^{(1)} = \lambda \underset{\mathbf{n}}{\sum} \left(G_{\mathbf{n}}-q_{\mathbf{n}}\right)^{2} + \mu \underset{\mathbf{n},k}{\sum} \left(L_{z,\mathbf{n}}^{k}\right)^{2} + 2\Omega \underset{diag}{\sum} \left(c^{\dagger}_{\mathbf{n}}d_{\mathbf{n'}} + h.c. \right) \equiv H_{G} + H_{E} + H_{R}
\end{equation}
\end{widetext}
which is similar to the Hamiltonian obtained by us previously \cite{Zohar2011} \footnote{The correspondence is to the Hamiltonians $H_0+H_R$ in the second and fourth paragraphs of the third page of \cite{Zohar2011}, as follows: $L_z \sim \delta$; $c,d \sim \tilde a, \tilde b$; $G - q \sim G$.} and from which, due to the scale hierarchy $\lambda \gg \mu , \Omega$, we shall obtain an effective Hamiltonian as in \cite{Zohar2011}. The constraint will be $H_{G}$ (Gauss's law), and its ground sector contains the states of relevance for us. $H_{E}$ commutes with it and hence becomes the first order of the effective Hamiltonian, and from $H_{R}$ we get two conributions: one is  the gauge invariant plaquette term $H_{B}=-\frac{8\Omega^{2}}{\lambda}\underset{m,n}{\sum}
\left(c_{\mathbf{n + \hat 2}}^{\dagger}d_{\mathbf{n}}c_{\mathbf{n}}^{\dagger}d_{\mathbf{n + \hat 1}} + H.C.\right)$. The other one is due to the finiteness of the angular momentum representation matrices, but it is diagonal (and hence gauge invariant) and thus introduces a negligible first order correction to the energy but does not change the zeroth order ground state (the flux tube): $H_{B}'=-\frac{2\Omega^{2}}{\lambda}\underset{diag}{\sum}\left( \left|+\right\rangle \left\langle+\right| + \left|0\right\rangle \left\langle0\right| \right)_{\mathbf{n}} {\otimes} \left( \left|0\right\rangle  \left\langle 0\right| + \left|-\right\rangle \left\langle-\right|\right)_{\mathbf{n'}} $. Note that as $l \rightarrow \infty$ (in a truncated Kogut-Susskind theory),  these terms approach identity matrices and act as an ignorable costant energy, and hence this term did not appear in the infinite case. An example for the emergence of gauge invariance as the constraint gets stronger is illustrated in figure \ref{eff}.

Next we perform the change of variables \cite{Zohar2011}: $L_{z,\mathbf{n}}^{k}\rightarrow(-1)^{n_1+n_2}L_{z,\mathbf{n}}^{k}$, $\phi_{\mathbf{n}}^{k}\rightarrow(-1)^{n_1+n_2}\phi_{\mathbf{n}}^{k}$ and
$Q_{\mathbf{n}}\equiv(-1)^{n_1+n_2}q_{\mathbf{n}}$ (which also swaps the $L_{\pm}$ operators on odd vertices).
This change of signs is needed in order to get the correct signs for the Gauss's Law constraint and the interactions of the Spin-Gauge Hamiltonian.
Plugging it into the effective Hamiltonian we get
\begin{equation}
H_{eff}^{(2)} = \mu \underset{\mathbf{n},k}{\sum} \left(L_{z,\mathbf{n}}^{k}\right)^{2} + H_{B} + H_{B}'
\end{equation}
and if we rescale the energy, using $\alpha=\frac{2}{g^{2}}\mu=\frac{16\Omega^{2}g^{2}}{\lambda}$, we get the that $\alpha^{-1}H_{eff}^{(2)}$ is the $l=1$ Spin-Gauge Hamiltonian equation (\ref{SG}) plus the irrelevant term of $\alpha^{-1}H_{B}'$. Hence, because of the sign inversions between the Spin-Gauge and the simulating models, the manifestation of an electric flux-tube between two confined static charges, to zeroth order, will be a line of alternating $+,-$ states of the atoms along the links connecting the two charges (see figure \ref{lattice}c). Some specific suggestions for initial state preparation and possible measurements are presented in the supplementary material.

In this paper, we have presented a new method to simulate cQED using ultracold atoms in optical lattices. We believe that this method may be experimentally accessible in the near future. Although we have constructed a realization of $H_1$, which allows for simulation of confinement around the strong coupling limit, the rapid convergence of $H_l$ to the Kogut-Susskind model suggests an avenue towards the simulation of the nonperturbative effects of the weak coupling limit as well as phase transitions in 3+1 dimensions. It would be intriguing to study the inclusion of dynamic matter fields in the model, which would lead to a full simulation of cQED.

\emph{Acknowledgments.} The authors would like to thank L. Mazza for helpful discussions.
BR acknowledges the support of the Israel Science Foundation, the German-Israeli
Foundation, and the European Commission (PICC). IC is partially supported by the EU project
AQUTE.

\section{Supplementary Material}

\subsection{The superlattice structure}

In the paper, we have suggested a method for simulation of the $l=1$ spin-gauge hamiltonian, as a simulation model for cQED. The effective derivation of the gauge-invariant Hamiltonian relied on Hamiltonian (2) in the paper,
which is a generalization of the model of \cite{Mazza2010}, where in section IV and appendix A the method of obtaining the tunneling rates is explained in detail.

In our case, the system is 2d and thus the generalization should be explained. As in \cite{Mazza2010}, the different hopping rates are achieved using a suplerlattice structure, in which unoccupied auxiliary minima, which are coupled by lasers to the "regular" occupied atomic minima, are adiabatically eliminated forming the desired hopping processes.
\begin{figure}[h]
\includegraphics[scale=0.7]{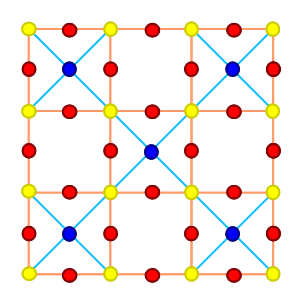}
\caption{\emph{The Superlattice Structure} A rotated plot of the optical lattice ($45^{0}$ degrees rotation from figure 1 in the paper). Yellow circles denote the occupied sites (which are on the links of the paper's model, which are the blue lines, and on the vertices here). Blue and Red circles denote the auxiliary sites of two different depths, through which tunneling processes between yellow sites  are generated.}
\end{figure}

In our model, the atomic minima of interest are located on the links of a spatial 2d lattice (see figure 2a in the paper). In figure 1 of the supplementary material, we present a rotated version of the lattice, which includes the auxiliary sites as well. As one can see in the figure, there are two types of auxiliary sites, with two different depths.
This can be achieved using an optical potential of the form

\begin{multline}
V\left(\mathbf{x}\right) = -V_0 \big[\cos ^2\left(2kx\right)\cos ^2\left(2ky\right) + \\
\cos ^2\left(kx\right)\cos ^2\left(2ky\right) + \\
\cos ^2\left(2kx\right)\cos ^2\left(ky\right) + \\
\cos ^2\left(\frac{kx}{2} - \frac{\pi}{4}\right)\cos ^2\left(\frac{ky}{2} - \frac{\pi}{4}\right) + \\
\sin ^2\left(\frac{kx}{2} - \frac{\pi}{4}\right)\sin ^2\left(\frac{ky}{2} - \frac{\pi}{4}\right) \big]
\end{multline}
 where $V_0, k > 0$ (see figure 2).

\begin{figure}[h]
\includegraphics[scale=0.35]{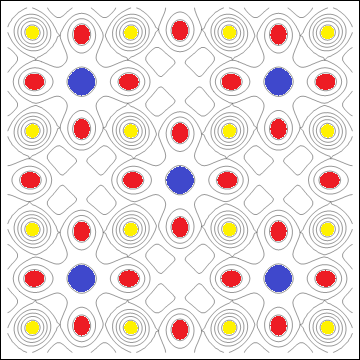}
\caption{A Contour Plot of the optical potential. The minima are colored according to figure 1. The yellow minima are the deepest.}
\end{figure}

To achieve the requested experimental scheme, one laser connects (along the blue lines in figure 1) the $m= \pm1$ levels of "real","yellow" sites and the same levels of the neighboring "blue" auxiliary sites, giving rise
 (using adiabatic elimination) to the (equal)
tunneling rates $t_\pm^s, t_\pm^d$ between "yellow" atoms. Another laser connects (along the red lines in figure 1) the $m=0$ levels of "yellow" sites with the same levels of the "red" sites. These lasers are responsible to $t_0^d$ in a similar manner. No other lasers are required, since we need $t_0^s=0$.

\subsection{State preparation and possible measurements}

Suppose, first, that the bosonic interactions ($H_B , H'_B$) are turned off ($\Omega = 0 \Leftrightarrow t_0^d = 0$) - i.e., the system is
in the strong coupling limit. The gauge field vacuum (when for every vertex $\mathbf{n}$, $q_{\mathbf{n}}=0$, does not contain any excited links: all the atoms are in their $m=0$ state.
This is the exact QED vacuum in the extreme strong limit, since no plaquette terms are present.
Thus the system has to be initially prepared in this state, which is an eigenstate of $H_G + H_E$ (and hence they do not need to be turned on adiabatically - if they were initially of, any way of turning them on can not change the state, but the Gauss's law constraint will be imposed).

After obtaining this state as a first stage, one can either do nothing or obtain some interesting initial states in the following methods; Most of them involve an adiabatic increase of $\Omega$, which is an adiabatic increase of $t_0^d$:
\begin{enumerate}
\item Remain in the strong limit; Do nothing else.
\item Obtain the QED vacuum (up to first order): increase $\Omega$ adiabatically, keeping $\frac{\Omega^2}{\lambda} \ll \mu$. This would effectively mean an adiabatic increase of $H_B$ (and $H'_B$), dressing the system to the desired state. Then, if $\Omega$ remains fixed after the adiabatic evolution, QED dynamics (to first order) will govern.
    One can use different $\Omega$s throughout the lattice, thus "focusing" the loops to certain areas (for example, to the center).
    In this case, the corrections to the strong limit vacuum can be calculated perturbatively. It can be shown using the theorem in the next section, that for $l=1$, the perturbative series of the vacuum states for the Kogut-Susskind and the Spin-Gauge theories are identical to first order; The first
    order correction is just a superposition of single excited plaquettes all over the lattice.
\item Obtain a "loop sea" by increasing $\Omega$ adiabatically again, but to $\frac{\Omega^2}{\lambda} \gg \mu$. Thus the system will be dressed to a state with lots loops, but with larger amplitudes. Then, when the state is ready, lower the value of $\Omega$ to obtain strong regime QED dynamics.
\item Create "Manually" zeroth order excitations: using single-addresing lasers \cite{Bakr2009,Weitenberg2011}, one can create closed flux-loops, or, if there are nonzero $q_\mathbf{n}$s, one can generate the required zeroth order
     flux-tubes, which are made of alternating signs atoms, as explained in the paper. Afterwards $H_B$ can be turned on adiabatically to obtain QED dynamics (note that in this case the highest accurate order is the zeroth order, as explained in the next section, but some corrections to it do appear).
\end{enumerate}

A qualitative glimpse on the structure of the state can be achieved  if the density of atoms of each species is measured and "frozen" (by switching off $\Omega \sim t_0^d$ after the measurement). The state collapses to a single element of the superposition, and flux tubes and loops be observed.
Being in the confining phase, only finite flux lines can be seen: either closed loops (single plaquettes, since this is the leading order contribution to the state) or flux tubes between static charges, if there are any. In the latter case the flux-tube prepared in zeroth order should remain, but with an additional loop somewhere around the lattice.     Although the probability for a particular loop is small, as a perturbative correction, if the system is large enough, the probability for a loop (somewhere around the lattice) becomes larger. This is unless the initial state was the "loop sea" - in that case, other interesting loop structures may be observed.

One should also note that the predictions given are based on perturbation theory considerations, and thus a realization of these measurements may serve as an interesting experimental check of them.

\subsection{Accuracy of $H_l$ eigenstates in perturbation theory}
\emph{Theorem.} Let $\left\{ N_{k}\right\} $ be operators with eigenvalues
$n_{k}\in\mathbb{Z}$ and eigenstates $\left|n_{k}\right\rangle $
satisfying $N_{k}\left|n_{k}\right\rangle =n_{k}\left|n_{k}\right\rangle $.
Let $\left\{ a_{k},a_{k}^{\dagger}\right\} $ be lowering and raising
operators, such that $a_{k}\left|n_{k}\right\rangle \propto\left|n_{k}-1\right\rangle ,a_{k}^{\dagger}\left|n_{k}\right\rangle \propto\left|n_{k}+1\right\rangle $.
Consider the Hamiltonian
\begin{equation}
H=H_{0}\left(\left\{ N_{k}\right\} \right)+\lambda V\left(\left\{ a_{k},a_{k}^{\dagger}\right\} \right)
\end{equation}
where $\lambda\ll1$, and the maximal power of any lowering\textbackslash{}raising
operator in $V$ is $m$, and a truncated version of the Hamiltonian,
\begin{equation}
\widetilde{H}=H_{0}\left(\left\{ \widetilde{N}_{k}\right\} \right)+\lambda V\left(\left\{ \widetilde{a}_{k},\widetilde{a}_{k}^{\dagger}\right\} \right)
\end{equation}
where the eigenvalues and eigenstates of $\left\{ \widetilde{N}_{k}\right\} $
satisfy $-l\leq n_{k}\leq l$ and $a_{k}\left|-l\right\rangle =0,a_{k}^{\dagger}\left|l\right\rangle =\left|0\right\rangle $.
Suppose that in $\left|G^{(0)}\right\rangle $, the ground state \footnote{Note that the derivation is true for any nondegenerate eigenstate, but the ground state is the relevant state for our purposes.} of
$H_{0}\left(\left\{ N_{k}\right\} \right)$, the highest eigenvalue
of an $\left\{ N_{k}\right\} $ operator (in an absolute value) is
$n_{0}$.

Expand the ground states $\left|G\right\rangle ,\left|\widetilde{G}\right\rangle $
of $H,\widetilde{H}$ respectively in time independent perturbation
theory,
\begin{equation}
\left|G\right\rangle =\left|G^{(0)}\right\rangle +\underset{n=1}{\overset{\infty}{\sum}}\lambda^{n}\left|G^{(n)}\right\rangle
\end{equation}
\begin{equation}
\left|\widetilde{G}\right\rangle =\left|\widetilde{G}^{(0)}\right\rangle +\underset{n=1}{\overset{\infty}{\sum}}\lambda^{n}\left|\widetilde{G}^{(n)}\right\rangle
\end{equation}
then for every $n\leq\frac{l-n_{0}}{m}$,
\begin{equation}
\left|G^{(n)}\right\rangle =\left|\widetilde{G}^{(n)}\right\rangle
\end{equation}

\emph{Proof}. Start with $n=0$. In order to have $\left|G^{(0)}\right\rangle =\left|\widetilde{G}^{(0)}\right\rangle $,
given that $\left|G^{(0)}\right\rangle $ contains the maximal eigenvalue
$n_{0}$, we need that $l\geq n_{0}$. Next, for $n=1$, the highest
possible maximal eigenvalue (abs. val.) is $n_{0}+m$, by applying
the raising\textbackslash{}lowering operators in a maximal power of
$m$. Thus one needs, for first order accuracy, that $l\geq n_{0}+m$.
Continuing to higher orders, we get that for $n$th order accuracy,
one must have $l\geq n_{0}+nm$, and hence, finally, $n\leq\frac{l-n_{0}}{m}$.
QED.

For the Hamiltonian of compact lattice QED, the
theorem applies to the strong coupling limit. There, $H_{0}$ is the
electric part, and $V$ is the magnetic part. Since the plaquettes
contain a single power of raising\textbackslash{}lowering operators,
$m=1$. In the absence of external charges, $n_{0}=0$, and thus $n\leq l$,
i.e., if the electric fields are truncated in a value of $l$, this
will be the highest accurate order. If there are, for example, two
opposite static charges connected by a flux tube, $n_{0}=1$, and
thus $n\leq l-1$, i.e., to be accurate to order $n$, the truncation
limit must be at least $n+1$.

For example, approximating the system with spin-1 spinors, one gets
accuracy to first order in the pure gauge case but only to the zeroth
trivial order in the case of external charges, for which an approximation
using $l=2$ is needed in order to get first order accuracy.

 For the sake of a complete discussion, one should note that in order to apply the theorem to our class of truncated operators, we should also consider the angular momentum nature of the spin-gauge model.
     In the $l=1$ case, although the ladder operators are not unitary, their nonvanishing matrix elements are identical and hence this problem can be disregarded. We shall coment that a "cure" for this problem (using some local counter terms) exists for $l=2$.

\bibliography{ref}

\end{document}